\documentclass[11pt]{article}

\usepackage{graphicx}
\usepackage{amsmath}
\usepackage[colorlinks=true,linkcolor=red,urlcolor=blue]{hyperref}
\usepackage{setspace}
\onehalfspace

\begin{document}
\begin{center}
{\huge On two optomechanical effects\\of laser-induced electrostriction\\in dielectric liquids\footnote{Copyright (2014) The Japan Society of Applied Physics.}}

\vspace{16pt}
{\sc Ardian B. Gojani\footnote{Institute of Fluid Science, Tohoku University, ardian582@gmail.com.} \& Rasim Bejtullahu\footnote{Department of Physics, Faculty of Mathematical and Natural Sciences, University of Prishtina.} \& Shigeru Obayashi\footnote{Institute of Fluid Science, Tohoku University.}}

\vspace{16pt}
\end{center}

\begin{spacing}{1}
\noindent \textbf{Abstract:} This paper presents electrostriction from the phenomenological perspective, and gives details on two mechanical effects arising from laser-matter interaction. Electrostriction is the tendency of materials to compress in the presence of a varying electric field. In this paper, the investigated materials are polar and nonpolar dielectric liquids. It is stressed that the dominant factor is the time evolution of the laser pulse, which causes tensile stresses and acoustic waves. The study is supported by experimental realization of electrostriction, which can be detected only at favourable conditions (observed in water, but not in castor oil). This study will shed light in developing measurement techniques (e.g., laser-induced grating spectroscopy) and in explaining the onset of cavities and laser-induced liquid breakdown.
\end{spacing}

\section{Introduction}

Lasers have become a tool of choice in many applications, mainly due to their directionality, monochromaticity, and coherence, which allow achievement of enormous energy densities in matter through which the laser is propagating. In addition, their short pulse duration gives rise to phenomena that would not occur otherwise.

This paper discusses electrostriction and its role in the initiation of two effects: induction of tensile stresses and generation of acoustic waves. The former is of paramount relevance for understanding mechanisms of cavity formation and initiation of laser-induced liquid breakdown,\cite{Kao, Smith} while the latter is an important component of optoacoustic based measurement techniques.\cite{Sig} Electrostriction plays an important role also in the manipulation of microdroplets and liquid surfaces by laser pulses \cite{EB_OL12, E_PF12}. These effects are not limited only to laser radiation, but can be achieved also by the discharge of electrodes \cite{SP_JAP13, PS5_JPD14, S_PSCT13}.

Initially, the theoretical framework for calculation of electrostriction is presented, and then the conditions for electrostriction-induced cavity formation are given. Experimental detection of electrostriction is a very difficult task, but the effect in water is noticeable. This is shown by comparing the laser-induced grating spectroscopy signals for water and castor oil.

\section{Theory of electrostriction}

Propagation of an electromagnetic wave through a dielectric medium causes the displacement of charge distribution, thereby inducing an electric dipole moment (causing polarization). In a first (linear) approximation, the electro-optical properties of the medium, such as refractive index, permittivity, susceptibility, polarizability, etc., can be considered independent of the electric field. But, if the strength of the applied electric field is large - of the order of $10^6$ V/m - these properties become dynamic. As a result, some phenomena, including  second harmonic generation, Kerr effect, self-focusing, thermalization, etc, can not be neglected. Among these is electrostriction. According to Boyd \cite{Boy_NO_08}, electrostriction is the tendency of materials to become compressed in the presence of an electric field, while, paraphrasing Landau and Lifshitz \cite{LL_ECM_984}, application of an external electric field on a medium causes a redistribution of mass density such that the free energy of the system is minimized, yielding a deformation of the medium; this deformation is electrostriction.

Although electrostriction is a phenomenon that takes place at the molecular level, it can be described by macroscopic properties of the medium. Thus, electrostriction has the nature of an excess pressure and its value can be calculated based on the Helmholtz-Lippmann equation: \cite{Gar_CEN_12}
\begin{equation}\label{HL}
p_{es} = -\frac{\epsilon_0 E^2}{2} \left(\rho_0 \frac{\partial \epsilon}{\partial \rho}\right)_T,
\end{equation}
where $E^2$ is the intensity of the electric field, $\epsilon$ is the relative permittivity of the medium, $\epsilon_0 = 8.854 \times 10^{-12}$ F/m is the vacuum permittivity, $\rho$ is the mass density of the medium, and $T$ is the temperature. Consequently, electrostriction acts as a force density
\begin{equation}
\mathbf{f} = - \nabla p_{es}.
\end{equation}

Based on Eq. (\ref{HL}), one cause of electrostriction is the change of permittivity of the medium due to the change of density, and can reach appreciable values only at high electric field intensities. This is so because of the low value of permittivity of vacuum. One method of reaching these high values is by focusing laser beams. Another cause is the sharp spatial and temporal change of the electric field itself, as it will be argued shortly.

In other words, the electrostrictive pressure field in an isotropic dielectric liquid through which a pulsed laser beam is propagating, is a product of two factors (consists of two components): the gradient of laser irradiance ($E^2$), and the change of the permittivity with density ($\rho_0 \partial \epsilon/\partial \rho$). In order to calculate the former, a model of the electric field of the laser has to be determined. Approximating to a spatially cylindrical beam of unit radius (collimated beam along the Rayleigh range), in other words, looking only on the spatial distribution of the irradiance in a cross-sectional plane normal to the direction of the propagation, the irradiation of a typical Gaussian laser beam (fundamental transverse mode TEM$_{00}$) is
\begin{equation}\label{laser}
I = E^2(r,t) = C' \exp(-2r^2) \,\tau(t),
\end{equation}
where $C'$ is a constant of the laser properties, $r$ is the scaled radial coordinate from the center of the beam, and $\tau(t)$ is the temporal envelope of the laser pulse. The scaling is done with respect to the waist size of the laser beam ($w_0=1$ unit of length, typically 10 $\mu$m), where
\begin{equation*}
r^2=\frac{x^2+y^2}{w_0^2}.
\end{equation*}
In subsequent calculations, the nondimensionality of time is achieved by introducing the scaled time
\begin{equation*}
\theta=\sqrt{2} \,\frac{c_s}{w_0} \,t.
\end{equation*}
In this case, $c_s$ is the speed of sound, and the time duration of the laser pulse $t$ is scaled based on the time it takes an acoustic wave to traverse the laser waist size. Typical values for $t$ and $c_s$ are 10 ns and 1.5 km/s, respectively.

When considering the polarity of liquids, there exists a distinction between nonpolar dielectrics, e.g., castor oil, and polar dielectrics, e.g. water, which is also reflected in the functional dependency of permittivity on the density \cite{Ush_IBL_07}. For nonpolar liquids, this relationship is described by the Clausius-Mossotti equation,\cite{BW} from where 
\begin{equation}\label{nonpolar}
\rho_0 \frac{\partial \epsilon}{\partial \rho} = \frac{(\epsilon-1)(\epsilon+2)}{3}.
\end{equation}
For polar liquids, the Onsager equation is applicable,\cite{Frenkel} giving
\begin{equation}\label{polar}
\rho_0 \frac{\partial \epsilon}{\partial \rho} \approx 1.5\epsilon.
\end{equation}

Employing permittivity (cf. refractive index through the Lorentz-Lorenz equation) is more convenient, because the frequency of the field corresponds to the temporal envelope of the laser pulse (for a nanosecond laser pulse, this frequency is $\sim 10^{9}$ Hz), and not to the frequency of light ($\sim 10^{14}$ Hz). For the already mentioned dielectrics in standard conditions \cite{Lid_CRC}, $\epsilon_{\text{water}} \approx 80$ and $\epsilon_{\text{castor oil}} \approx 4.7$; that is, electrostrictive pressure in water is an order or magnitude larger than in castor oil.

Substitution of Eqs. (\ref{laser}) and (\ref{nonpolar}) or (\ref{polar}) into the equation for electrostrictive pressure (\ref{HL}), gives
\begin{equation}\label{Pes}
p_{es}(r,t) = - C \exp(-2r^2) \,\tau(t),
\end{equation}
where $C$ collects all the constants of the equation, specific to a particular experimental arrangement (laser and dielectric liquid properties).

\section{Results and discussion}

\subsection{Calculation of electrostrictive pressure}

Since the term that describes the external source of the force is available, the calculation of the pressure field can be done relatively easy. Several different approaches are presented in the literature.

One is to consider the electrostrictive pressure as a source term into the acoustic wave equation \cite{She_PL966}
\begin{equation}\label{acoustic}
\nabla^2 p - \frac{1}{c_s^2}\frac{\partial^2 p}{\partial t^2} +\frac{2\Gamma_A}{c_s^2}\frac{\partial p}{\partial t} = \nabla \cdot \mathbf{f},
\end{equation}
where $\Gamma_A$ is the acoustic damping, and 
\begin{equation}\label{csound}
c_s^2 = \left(\frac{\partial p}{\partial \rho}\right)_S
\end{equation}
is the speed of sound (the speed of the propagation of a disturbance $p$ and $\rho$). Obviously this can be done only if the equation of state for the liquid is provided, and many experimentally measured parameters for liquids are available \cite{GOT_SH_09}.

Another approach is to directly insert the equation of electrostrictive pressure (\ref{Pes}) into the standard system of equations for the conservation of mass and momentum (Euler equations), and solve those numerically, as it has been pursued by Shneider and Pekker \cite{SP_PRE13}.

A generalization is provided by Ellingsen and Brevik \cite{EB_efp_PF11}, who consider the flow to be potential and cylindrically symmetric, and express the velocity by the potential as $\mathbf{v} = \nabla \Phi$, for which the governing equation is
\begin{equation}\label{phi}
\nabla^2 \Phi - \frac{1}{c_s^2}\frac{\partial^2 \Phi}{\partial t^2} = - \frac{1}{\rho_0 c_s^2}\frac{\partial p_{es}}{\partial t}.
\end{equation}
Substitution of Eq. (\ref{phi}) into linearized Euler equations gives for pressure
\begin{equation}\label{pressure}
p = -\rho \frac{\partial \Phi}{\partial t} + p_{es}.
\end{equation}
If the function of the laser time pulse $\tau(t)$ is given, this equation can be solved and the pressure can be evaluated.

\begin{figure}
\begin{center}
\includegraphics{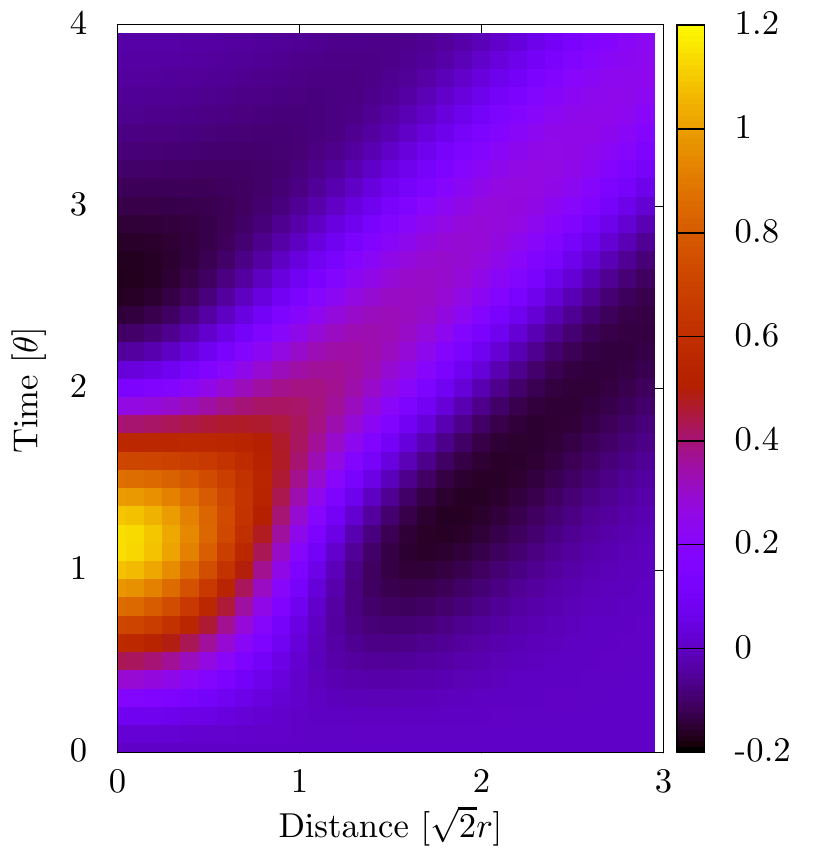}
\end{center}
\caption{Electrostrictive pressure field map, its evolution and distribution from a laser with a unit rectangular pulse.}\label{map}
\end{figure}

A simple example of a laser pulse is the temporal unit step function, which has the irradiance
\begin{equation}
I = C \exp\left(-2r^2\right) \left[\Theta(t)-\Theta(t-1)\right].
\end{equation}
Here $\Theta(t)$ is the Heaviside step function. The pressure for this type of laser pulse is calculated using the QAGS routine of the QUADPACK library,\cite{Eins_ASC05} and the result is shown in Fig. \ref{map}. The magnitude bar shows the ratio of the electrostrictive pressure to the equilibrium pressure, and, as it can be seen, the overall pressure varies up to $\pm 20\%$. Pressure varies in the case of sudden changes, it is built and reaches equilibrium in about 3-4 longer times than the scaled time interval (the time it takes an acoustic wave to pass the laser beam).

Two observations can be made: (i) electrostriction generates an acoustic wave, and (ii) electrostriction induces tensile stresses (negative pressures) in liquids. While the electrostriction induced acoustic wave will be present at all conditions, tensile stresses will not. This will be illustrated by looking at the pressure values along the central line of the laser beam, i.e., for $r=0$.

\subsection{Electrostriction and tensile stresses}

Along this line the pressure is \cite{EB_efp_PF11}
\begin{equation}\label{pline}
p=\tau(\theta) - \left[\tau'(0)F(\theta)+\int_0^{\theta} \tau''(\theta-s)F(s) ds\right],
\end{equation}
where
\begin{equation}
F(s) = \exp(-x^2) \int_0^x \exp(s^2) \, ds
\end{equation}
is the Dawson integral.\cite{NIST_math} If at some interval the sum in the second term of the equation for pressure is larger than the pulse function, the pressure will have negative values. Since Dawson integral is positive for all values of the argument, then the pressure is influenced by the rate change of the pulse. For example, pulses that increase linearly or quadratically do not produce negative pressure, no matter how intense they are. A similar behavior is observed in the case of a Gaussian pulse duration.

\begin{figure}[t]
\begin{center}
\includegraphics[width=\linewidth]{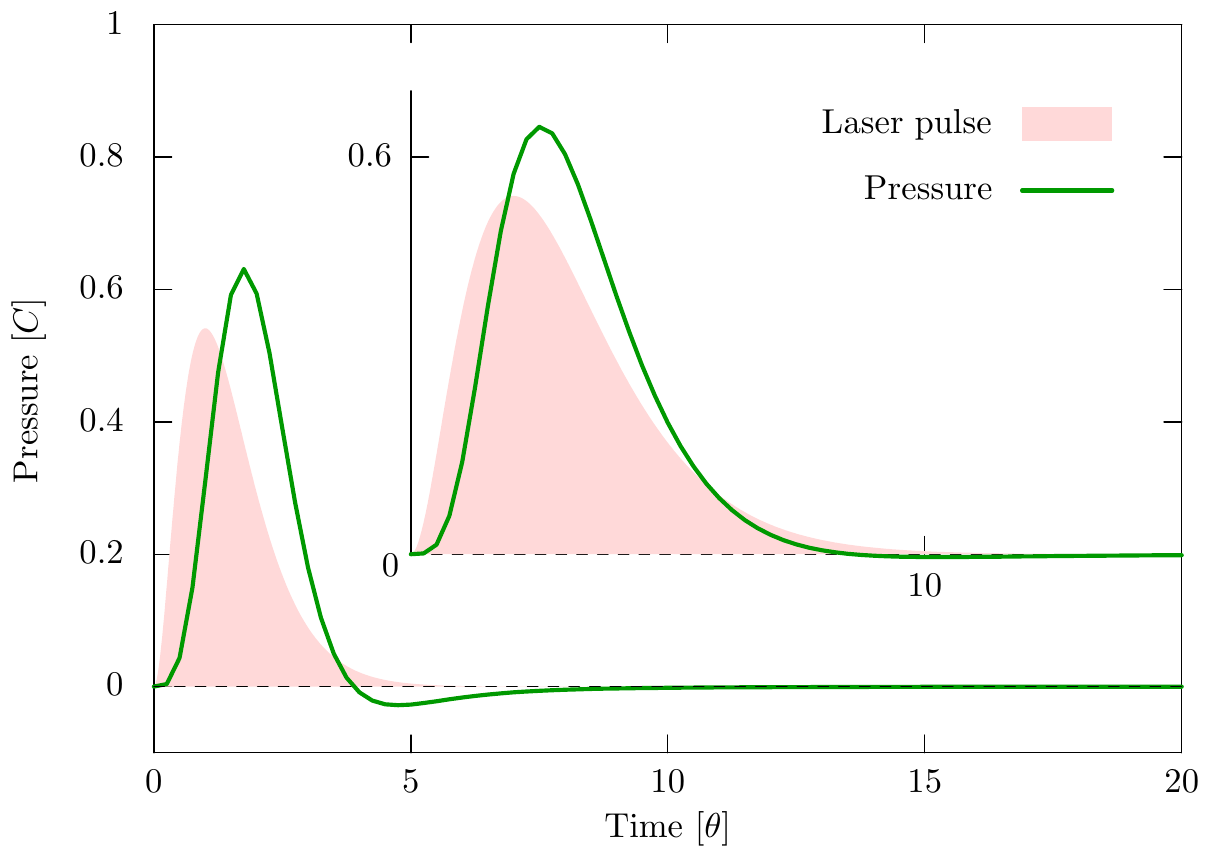}
\end{center}
\caption{Pressure evolution (green line) caused by the laser pulse (red area) ${\tau(t)\propto t^2 \exp(-t)}$. Note that laser energies are one (lower graph) and two units (the graph shifted for clarity). Pressure unit is $C$ as defined in equation \ref{Pes}.}\label{assym}
\end{figure}

An interesting observation can be made for pulses that are not symmetric, which are shown in Fig. \ref{assym}. These two laser pulses are given by the relation
\begin{equation}
\tau(t) \propto t^2 \exp{-t}
\end{equation}
and they differ only by the pulse duration (one is double the other). Again, two observations can be made:
\begin{itemize}
\item the time to reach equilibrium is much longer, and this is due to the \emph{softness} of the falling pulse, and
\item if the pulse is long enough, negative pressures will not be achieved, and this is due to the motion of the liquid.
\end{itemize}

Every liquid is characterized by a critical tension (a negative pressure) beyond which the fluid will rupture and cavitation will be formed \cite{Bre_CBD_995}. One possibility for determining the threshold values for this pressure is by considering the nucleation theory, according to which
\begin{equation}\label{dpc}
\Delta p_c = \left(\frac{16\pi}{3}\frac{S^3}{W_{cr}}\right)^{1/2},
\end{equation}
where $p_c$ is the tensile strength of the liquid, $S$ is the surface tension, and $W_{cr}$ is the energy that must be deposited to form the bubble (the work of bubble formation) and essentially is determined experimentally. 

Conditions obtained from equations (\ref{pline}) and (\ref{dpc}), namely
\begin{equation}
\tau(\theta) < \tau'(0)F(\theta)+\int_0^{\theta} \tau''(\theta-s)F(s)\,ds,
\end{equation}
\begin{equation}
|p| > |\Delta p_c|
\end{equation}
provide the necessary and sufficient conditions for inducing tensile stresses. This analysis shows that laser-induced electrostriction plays a role in cavitation formation, but the key parameter being the rise-fall time of the pulse (high pulse energy, for example, being insufficient for cavity formation in cases of \emph{smooth} pulses). 

\subsection{Experimental observation of electrostriction through laser-induced dynamic gratings}

\begin{figure}
\begin{center}
\includegraphics[width=\linewidth]{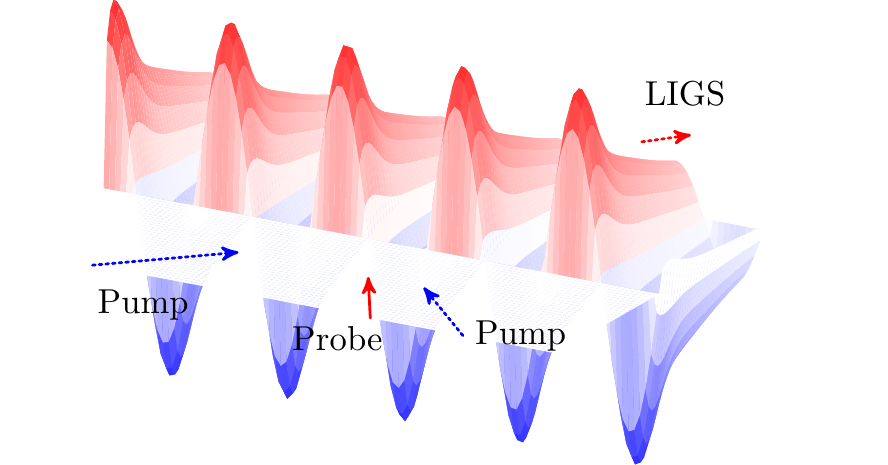}
\end{center}
\caption{Generation of dynamic gratings, where the electrostriction effect may be recorded. Pump beams generate the grating, probe beam is directed under the Bragg angle, while the LIGS signal arises due to the diffraction of the probe beam.}\label{ligs}
\end{figure}

Observation of electrostriction induced acoustic waves can be achieved through Brillouin scattering, i.e. scattering of light from sound waves. This is realized in laser-induced grating spectroscopy (LIGS) measurements \cite{NM4_JAP82, EGP_lidg_S986, C_lita_OE995}, as illustrated in Fig. \ref{ligs}. Details of the experiment are presented elsewhere \cite{G5_SPIE04}, but as a brief reminder, the principle of the technique is as follows: an Nd:YAG laser beam (wavelength 532 nm, pulse 10 ns, energy 60 mJ) is divided into two equal parts that are made to interfere in a liquid, producing an optical grating. The interference grating can be modeled by the change of the refractive index, which is expressed by
\begin{equation}
\Delta n \propto \exp(-\Gamma_T t) +\exp(-\Gamma_A t) \cos(\nu t).
\end{equation}
Here, $\Gamma_T$ is the thermal grating decay rate determined by thermal diffusion, $\Gamma_A$ is the acoustic damping coefficient, and $\nu$ is the Brillouin frequency. Two most important mechanisms that contribute to the creation of this grating are thermalization, which essentially is the heating of the medium through a sequence of light absorption and molecular collision processes, and electrostriction. When a probing beam irradiates the grating under the Bragg angle, diffraction takes place. In the experiment, this was achieved by using an Ar$^+$ laser (wavelength 488 nm, continuous beam, with power 0.4 W). The diffracted beam can be recorded, revealing the state of the grating.

\begin{figure}[t]
\begin{center}
\includegraphics[width=\linewidth]{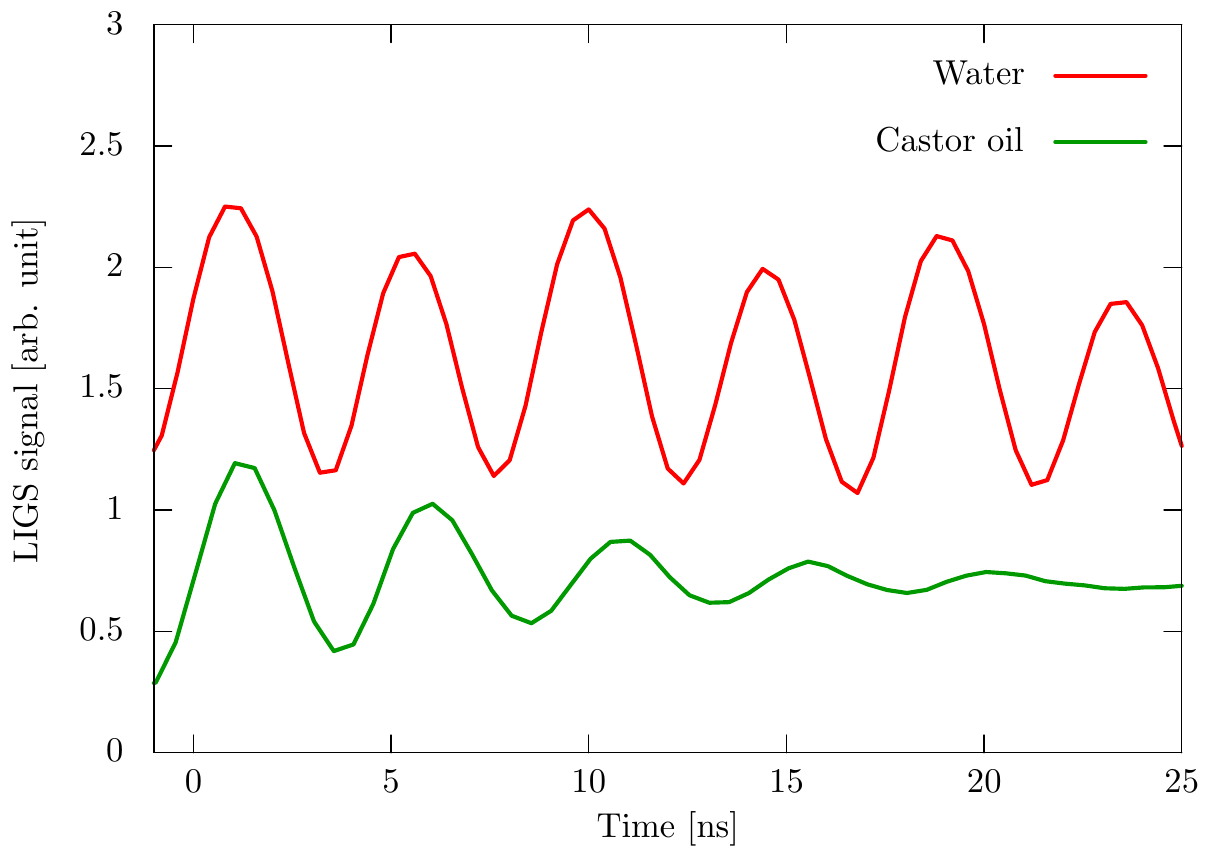}
\end{center}
\caption{LIGS signals in water and castor oil. Signal in water clearly shows contributions from electrostriction and thermalization, while the signal in castor oil does not record the contribution of the former.}\label{lita}
\end{figure}

The results from a LIGS experiment performed in water and castor oil are shown in figure \ref{lita}. The differences between signals are mainly due to thermal diffusivity and acoustic damping coefficient of each liquid, which can be calculated by measuring the normalized difference in height of the LIGS signal (the ratio of the decrease of the local maxima) \cite{B_IJT995, FD4_JCED06}. Such a measurement is not performed here because of the small number of recorded oscillations, consequently, large measurement uncertainties. It should be noted that the total increase of temperature of the irradiated medium does not exceed a fraction of a degree. 

\begin{figure}[t]
\begin{center}
\includegraphics[width=\linewidth]{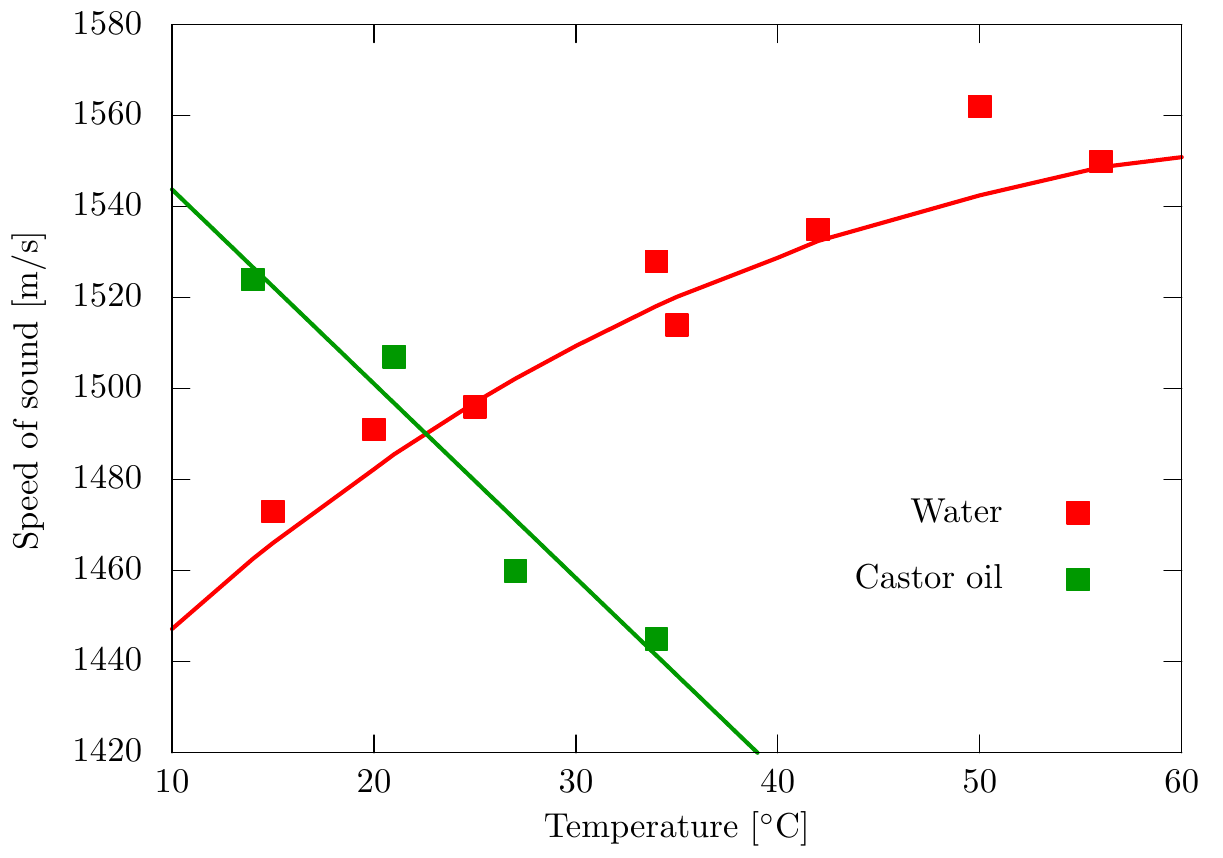}
\end{center}
\caption{LIGS measurements are applied for finding the relationship between the speed of sound and temperature in liquids.}\label{temp}
\end{figure}

But, what is important for the present theme, it can be clearly seen that the signal from water shows two frequencies, corresponding to thermalization and electrostriction, while for castor oil the signal due to electrostriction is so weak that it is not detectable (the effect for water is 10 times larger). These results can be applied in determining the relationship between the speed of sound and the temperature \cite{AD5_smt_AO05}, as shown in Fig. \ref{temp}.

Another possibility for observing electrostriction is by schlieren. This has been achieved for electric fields produced by high voltage electrodes, \cite{HH_PPS962} but not for laser induced electrostriction. It should be noted that visualization requires different treatment in regards to the chemical polarity of the liquids. For polar liquids, the nonlinear optical effects will be dominated by the Kerr effect, thus birefringence will be present in the images, while for nonpolar liquids, the dominant effect - although it is small - would be due to electrostriction.

\section{Conclusions}

In conclusion, electrostriction plays an important role in the stability and establishment of equilibrium in liquids, under the condition that the pulses that induce electrostriction are \emph{sudden} (with vary fast switch-on or off times). The electric field from a laser propagating through a liquid generates a pressure wave that in turn causes acoustic waves and tensile stresses. Since laser-matter interaction is a complex phenomenon, electrostriction is accompanied by several other nonlinear optical effects, which in many cases dominate. Comparison of polar and nonpolar dielectric liquids shows that electrostriction pressure in polar liquids is an order of magnitude larger, which allowed for its detection by LIGS. Electrostriction pressure and the subsequent disturbances can be used for the measurement of different liquid properties, and an example of temperature vs speed of sound measurement is given.

Considering interests in ultrashort laser-matter interactions, either for material processing, either for bioengineering applications, this paper put forward the notion that electrostriction may play an important role and should be taken into account during analysis.  

\hfill

\textbf{Acknowledgment}

ABG would like to express his thanks to Professor Emeritus Kazuyoshi Takayama, for his support.

\end{document}